\documentstyle[epsf]{l-aa}

\newcommand{\ep}{\epsilon}
\newcommand{\gag}{\gamma_{\scriptscriptstyle \rm g}}
\newcommand{\gac}{\gamma_{\scriptscriptstyle \rm c}}
\newcommand{\per}{\scriptscriptstyle \bot}
\newcommand{\md}{{\rm d}}
\newcommand{\rgp}{r_{{\rm g} \per}}

\newcommand{\tauesc}{\tau_{\rm\scriptscriptstyle esc}}
\newcommand{\kappaB}{\kappa_{\rm\scriptscriptstyle B}}
\newcommand{\boltz}{k_{\rm\scriptscriptstyle B}}
\newcommand{\nuesc}{\nu_{\rm\scriptscriptstyle esc}}
\newcommand{\fmax}{f_{\scriptscriptstyle\rm M}}
\newcommand{\gmax}{g_{\scriptscriptstyle\rm M}}
\newcommand{\gcr}{g_{\scriptscriptstyle\rm CR}}
\newcommand{\ncr}{n_{\scriptscriptstyle\rm CR}}
\newcommand{\fcr}{f_{\scriptscriptstyle\rm CR}}
\newcommand{\gcrt}{\tilde{g}_{\scriptscriptstyle\rm CR}}
\newcommand{\pinj}{p_{\scriptscriptstyle\rm inj}}
\newcommand{\pth}{p_{\scriptscriptstyle\rm th}}
\newcommand{\ca}{c_{\scriptscriptstyle\rm A}}
\newcommand{\eqb}{\begin{eqnarray}}
\newcommand{\eqe}{\end{eqnarray}}
\begin{document}
\thesaurus{12(02.01.1; 02.08.1; 02.19.1; 03.13.4; 09.03.2)}
\title{Time dependent cosmic--ray shock acceleration with 
self--consistent injection}
\author{U.D.J. Gieseler\inst{1}\thanks{{\it Present address:\/} 
     Universit\"at Siegen, Fachbereich Physik, 57068 Siegen, Germany} 
 \and T.W. Jones\inst{1} \and Hyesung Kang\inst{2}}
 \institute{ University of Minnesota, Department of Astronomy,
        116 Church St. S.E., Minneapolis, MN 55455, U.S.A. \and
   Department of Earth Sciences, Pusan National University,
       Pusan 609-735, Korea}
\offprints{ug@nesa1.uni-siegen.de}
\date{Received 5 September 2000; accepted 5 October 2000}
\maketitle
\begin{abstract}
One of the key questions to understanding the efficiency of diffusive 
shock acceleration of the cosmic rays (CRs) is the injection process 
from thermal particles. 
A self--consistent injection model based on the interactions of the 
suprathermal particles with self--generated magneto--hydrodynamic waves
has been developed recently by Malkov (\cite{Malk98}). 
By adopting this analytic solution, a numerical treatment of the
plasma--physical injection model at a strong quasi--parallel shock 
has been devised and incorporated into the combined gas dynamics and 
the CR diffusion--convection code. 
In order to investigate self--consistently the injection and acceleration 
efficiencies,
we have applied this code to the CR modified shocks of both high 
and low Mach numbers ($M=30$ and $M=2.24$) with a Bohm type diffusion
model. 
Both simulations have been carried out until the maximum momentum 
$(p_{\rm max}/m_{\rm p}c) \sim 1$ is achieved to illustrate early evolution
of a Bohm type diffusion.
We find the injection process is self--regulated in such a way 
that the injection rate reaches and stays at a nearly stable value after 
quick initial adjustment. 
For both shocks about $10^{-3}$ of the incoming thermal 
particles are injected into the CRs.
For the weak shock, the shock has reached a steady state within our 
integration time and $\sim 10\%$ of the total available shock energy 
is transfered into the CR energy density.
The strong shock has achieved a higher acceleration efficiency
of $\sim 20 \%$ by the end of our simulation, 
but has not yet reached a steady--state.
With such efficiencies shocks do not become CR--dominated 
or smoothed completely during the early stages when the particles are
only mildly relativistic. Later, as the CR pressure becomes
dominated by highly relativistic particles that situation should change,
but is difficult to compute, since the maximum CR momentum increases
approximately linearly with time for this model. In the near future we
intend to extend such shock simulations as these to include much
higher CR momenta using an adaptive mesh refinement technique currently
under development.

\end{abstract}
\section{Introduction}
\label{intro}
The non--thermal energy distributions of cosmic ray ions or source
distributions of electrons emitting synchrotron radiation in various 
astrophysical objects are commonly described as produced by the first 
order Fermi acceleration process  at shocks (for reviews see 
Drury~\cite{Drur83}; Blandford \& Eichler~\cite{BlEi87}; 
Kirk et al.~\cite{KiMePr94}). 

When particles diffuse\footnote{In general, also 
anomalous transport like {\em sub}--diffusion or {\em super}--diffusion
can be realized.}
off the moving scattering centers
in a region divided by a velocity discontinuity (shock),
these particles can be accelerated if their mean free paths exceed the shock 
thickness. The relative momentum gain for a cycle
of two crossings of the shock is then proportional to the velocity
difference across the shock, i.e.~of first order
with respect to the shock velocity (Bell~\cite{Bell78}). 
In astrophysical collisionless plasmas an electro--magnetic field must be 
present to change the energy of particles. Waves or irregularities in this
field provide particle scattering, which leads to diffusion.
Consider a shock with velocity $u_{\rm s}>0$ propagating into a plasma at 
rest with density $\rho$ and with a homogeneous magnetic field $B_0$ in 
the direction of the shock normal.
The plasma is compressed to the density $\rho_{\rm d}$ by the shock, and 
flows downstream with the velocity $u_{\rm d}=u_{\rm s}(1-1/r)$, where
$r= \rho_{\rm d}/\rho$ is the
compression ratio. 
Particles with the mean downstream velocity 
$\langle v \rangle =u_{\rm d}$ cannot cross the shock from downstream to 
upstream, because $\langle v \rangle < u_{\rm s}$.
In addition the shock may not be a discontinuity for a particle at this energy,
because the gyro radius of the thermal particles is of the order of the shock
thickness, leading to adiabatic energy change while crossing the shock.
Because the plasma is also heated downstream of the shock, 
some supra--thermal particles in the high energy tail of the Maxwellian
velocity distribution may gain energies and
have velocities that allow them to re--cross the shock. 
In a homogeneous magnetic field, due to lack of scattering centers
these particles would escape upstream without returning to
the shock. Then, the acceleration mechanism would not apply.
However, the population of particles that can move
upstream provide a seed particle beam, which generates Alfv\'en waves 
responsible for scattering and, therefore, diffusion, which is an essential 
element of the first order Fermi acceleration process. 

The problem of particle acceleration from thermal energies up to relativistic
particle energies is highly non--linear, as first pointed out by 
Eichler~(\cite{Eich79}). First, the energy transferred from the
bulk of the plasma to the sub--population of accelerated particles can change 
the thermodynamic properties of the plasma like the temperature and density. 
In addition, the accelerated particles provide their own pressure in the 
system, which, since it differs from the thermal pressure, modifies the 
velocity structure of the shock transition. 
Second, the waves generated by particles escaping upstream determine the 
transport properties of the plasma, and, therefore, regulate this wave 
generating escape itself. The manner in which the wave--particle interactions
control the fraction of plasma particles that can escape upstream to 
participate in the Fermi process is commonly called {\em injection}.
This is a basic aspect of the plasma of collisionless shocks and is itself
highly non--linear. This injection problem is fundamentally related to the
question of the efficiency of particle acceleration at shocks by the
Fermi process.

Different numerical methods have been used to treat the injection problem of 
CR modified shocks. 
In Monte--Carlo simulations of non--linear particle acceleration the details
of a posited scattering law provide an injection parameter, but one not
determined self--consistently from the particle wave interaction 
(e.g.~Ellison et al.~\cite{ElBaJo96}; Baring et al.~\cite{BaElReGrGo99}). 
In contrast to pure kinematical effects
from shock velocity, particle speed and inclination angle of magnetic
field and shock, the waves responsible for particle scattering depend on 
plasma properties like temperature and the beam strength of the wave 
generating particles itself. These coupled and time dependent effects are 
not easy to incorporate into a Monte--Carlo approach. Time dependent
Monte--Carlo simulations have been presented by Knerr et al.~(\cite{KnJoEl96}),
but still with a prescribed scattering law, as a parameterization of 
injection.

In the two--fluid approach the cosmic rays are treated as a diffusive
gas without following their momentum distribution. The energy transfer
into CRs in these models is based on a fraction of the upstream gas 
particles, that are instantaneously accelerated at the shock 
(Dorfi~\cite{Dorf90}), around the shock (Jones \& Kang~\cite{JoKa90}) 
or at velocity gradients (Zank et al.~\cite{ZaWeDo93}). In practical terms, 
because the shock is the most prominent velocity gradient in the system, 
these techniques are very closely related, as pointed out by 
Kang \& Jones~(\cite{KaJo95}).
Essentially the same parameterization is also used in
the numerical solution of the hydro--dynamical equations
coupled to the momentum dependent cosmic--ray transport equation 
(e.g.~Falle \& Giddings~\cite{FaGi87}; Kang \& Jones~\cite{KaJo91}; 
Berezhko et al.~\cite{BeYeKs94}). 

Kang \& Jones~(\cite{KaJo95}) 
used a numerical injection model with two essentially 
free parameters which describe boundaries in momentum at which
particles can be accelerated ($p_1 = c_1 \cdot \pth$
where $\pth$ is the peak momentum of the Maxwell distribution) 
and from which these contribute to 
the cosmic--ray pressure ($p_2 = c_2 \cdot \pth$).
Still, these momentum boundaries are free parameters,
which can be translated into a particle fraction of the upstream gas.
However, models that incorporate more details of the plasma physics
of the background plasma, and, therefore, the CRs at injection energies 
are really necessary to constrain the phase--space function and 
therefore determine these parameters. In this way we can incorporate
self--consistent plasma physical models into numerical simulations of
CR acceleration.

Such a plasma physical model based on non--linear interactions of particles 
with self--generated waves in a 
shocked plasma has been investigated numerically by solving the kinetic
equations of ions in a magnetic field, and treating the electrons as a 
background fluid (Quest~\cite{Ques88}). These simulations show that
ions can be scattered back and forth across the shock by self--generated
waves, and Quest~(\cite{Ques88}) also points out that these scattered
ions can provide a seed population of cosmic rays.

Recently, the kinetic equations of ions were solved analytically for
non--linear wave--fields near strong
parallel shocks by Malkov \& V\"olk (\cite{MaVo95}) and Malkov (\cite{Malk98}).
These authors were able to constrain the fraction of phase--space of the
background plasma that can be injected into the acceleration process as a 
result of the self--regulating interaction between wave generation and 
particle streaming. 
Here we incorporate this self--consistent analytical result in 
numerical solutions of the hydro--dynamical equations together with the 
cosmic--ray transport equation. 
Our simulations, therefore, provide the first time
dependent solution of the problem of modified shocks that includes
a self--consistent plasma physical injection model. 
This technique enables us to determine the 
level of shock modification and acceleration efficiency in an
evolving shock without a free 
parameter for the injection process (Gieseler et al.~\cite{GiJoKa99}).

We describe the plasma physical injection model in some more detail in 
Sect.~\ref{model} before we outline the coupled set of dynamical equations
of the plasma and cosmic rays in Sect.~\ref{dyn_model}, together with
details of the numerical method we used to solve these equations. 
The results are presented in Sect.~\ref{results} and Sect.~\ref{resultsweak}, 
consisting of the
time dependent evolution of the plasma properties, showing especially the
modification of the velocity profile and the momentum distributions
of thermal and relativistic cosmic--ray particles. In addition we present
our results in the form of a particle injection efficiency and 
an energy transfer efficiency at modified strong shocks.
\section{Injection Model}
\label{model}

\subsection{Wave--particle interaction at parallel shocks}
In supernova remnants (SNRs) the relative orientation of magnetic field and 
shock front can be very diverse, even in one single object. For example,
if a spherically symmetric shock front expands in a region of homogeneous 
magnetic field, the directions of the shock normal and the magnetic field  
change over the shock surface from parallel to perpendicular. 
For nearly perpendicular
shocks the acceleration process can be very fast and effective due to
reflections upstream of the shock (Naito \& Takahara~\cite{NaTa95}).
However, the velocity of the intersection point of
shock and magnetic field in highly oblique shocks can be close to light
velocity. 
This can suppress the injection efficiency of thermal particles, which
are effectively tied to magnetic field lines. 
That is because the velocity distribution function of thermal protons, which 
we shall assume to be Maxwellian, drops sharply towards high energies. 
This purely kinematical
effect has been investigated by Baring et al.~(\cite{BaElJo93}).
Therefore, regions of SNRs where quasi--parallel shocks exist, are likely 
to be where the most effective injection occurs. On the other hand, effective
acceleration, i.e.~short acceleration time scales and hard spectra,
may be realized in other parts of a SNR, where an oblique geometry
of magnetic field and shock normal is found.

\begin{figure}[t]
    \vspace{0cm}
    \begin{center} 
      \epsfxsize8.0cm 
      \mbox{\epsffile{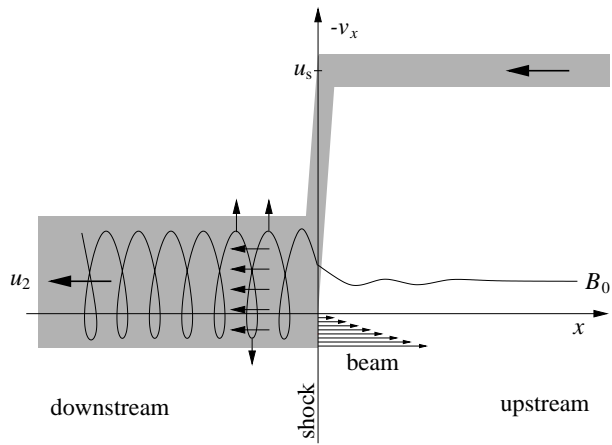}}
    \end{center}
    \vspace{0cm} 
    \protect\caption{Cartoon of the injection model in the 
shock--frame phase--space. Plasma is moving towards $-x$ into the shock 
with velocity $-u_{\rm s}$ and gets compressed,
heated and decelerated to the downstream velocity $-u_2$. Particles with
positive velocity can stream back to upstream along the magnetic field $B_0$.
These particles provide the beam, which generates the magnetic field 
waves.  The magnetic field wave is shown schematically in 
configuration space. The wave amplitude, frequency, and damping length 
is shown only qualitatively.}
\label{cartoon}
\end{figure}
 
For quasi--parallel shocks, where the shock propagates along the mean
magnetic field ($x$--direction), the transport properties along
the mean field direction are most important. We will 
assume this case, with the field $\vec{B}_0$ parallel to the shock normal.
The spatial diffusion of particles
is produced by magneto--hydrodynamic waves, which are in turn generated by 
particles streaming along the magnetic field, $\vec{B}_0$. 
We refer to Malkov~(\cite{Malk98}) for an extended analytical description
of the particle--wave interaction for low--momentum particles, 
and we describe here only the results
which are relevant for the implication of this model in our simulations
of the time dependent acceleration at modified shocks. When particles are
streaming along the magnetic field in the upstream direction, waves
are generated due to the ion cyclotron instability.
The resulting upstream magnetic field, which corresponds 
to a circularly polarized wave, can be written as 
\eqb
\vec{B}=B_0\vec{e}_x + B_{\per}(\vec{e}_y \cos k_0x -\vec{e}_z \sin k_0x).
\eqe
The amplitude $B_{\per}$ will be amplified downstream  of the shocks by a factor
$r=\rho_{\rm d}/\rho$. The downstream field can be described by a 
parameter $\ep$, for which, following Malkov, we assume 
$\ep := B_0/B_{\per}\ll 1$, in the case of strong shocks.
Note that the perpendicular component of the magnetic field leads effectively
to an alternating field downstream of the shock for particles
moving along the shock normal (see Fig.~\ref{cartoon}).

\subsection{Thermal leakage model}
The particles with a large enough gyro radius 
\eqb
\rgp= p\,c\,\sin\alpha/(eB_{\per}) > 1/ k_0
\eqe
can have an effective velocity with respect to the wave frame, i.e.~the 
downstream plasma would be transparent. Some of these
particles that are in the appropriate part of the phase space 
(depending on the shock speed) would be able to cross the shock from 
downstream to upstream. For the protons of the plasma, the resonance 
condition for the cyclotron generation of the Alfv\'en waves
gives $k_0 \langle v \rangle \approx \omega_0 = \omega_{\per}B_0/B_{\per}$, 
where the cyclotron frequency of protons is given by
 $\omega_{\per}=eB_{\per}/(m_{\rm p} c)$, and $\langle v \rangle$ is the mean 
downstream 
thermal velocity of the protons. We now have for the thermal protons
 $k_0 \rgp \approx \ep \ll 1$. 
{\it This means that most of the downstream thermal
protons would be confined by the wave, and only particles with higher velocity
in the tail of the Maxwellian distribution are able to leak through the shock.}
Ions with mass--to--charge ratio higher than protons have a proportionally 
larger gyro radius, so that the injection efficiency of protons would yield 
a lower limit for the less magnetized ions.
On the other hand, for thermal electrons a plasma with
such proton generated waves would have a reduced transparency due to 
the smaller gyro radius of the electrons. However, reflection
of electrons (and protons) off the shock could become efficient with
increasing wave amplitude and possibly aid in their injection 
(e.g.~Levinson \cite{Levi96}; McClements et al.~\cite{McDeBiKiDr97}).
In the following we will focus on the protons, which carry most of the energy 
and momentum of the plasma.

\subsection{Transparency function}
\begin{figure}[t]
    \vspace{-4cm}
    \begin{center} 
      \epsfxsize9.0cm 
      \mbox{\epsffile{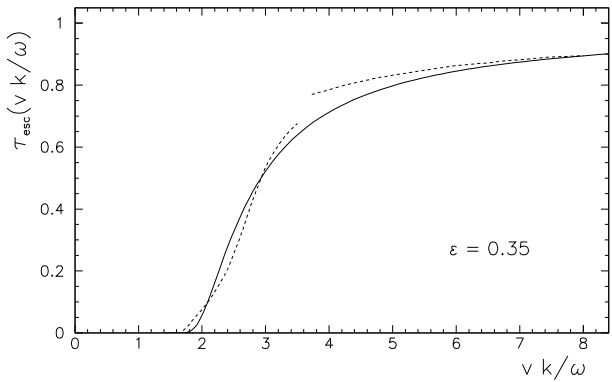}}
    \end{center}
    \vspace{-1cm} 
    \protect\caption{Transparency function Eq.~(3) vs.~the normalized 
particle velocity $\tilde{v}= vk_0/\omega_{\per}$ for
$\epsilon=0.35$ (solid line). 
Shown as a dashed line is the transparency function given in Fig.~2
of Malkov \& V\"olk~(1998).}
    \label{tau_plot}
\end{figure}
To find the part of the thermal distribution for which the magnetized plasma 
is transparent, and, which, therefore, forms the ``injection pool'', 
Malkov~(\cite{Malk98}) solves
analytically the equations of motion for protons in self--generated waves.
He finds a {\it transparency function} $\tauesc$, which expresses the
fraction of particles that are able to leak through the magnetic waves,
divided by the part of the phase space for which particles would be able to 
cross from downstream to upstream when no waves are present. 
For the adiabatic wave particle interaction the transparency
function  is given by 
Malkov~(\cite{Malk98}) Eq.~(33), with $\tauesc=2\,\nuesc/(1-u_2/v)$, 
where $v$ is the particle velocity $v$ and $u_2=u_{\rm s}/r$ is the velocity 
of the shock in the downstream plasma frame.
Here $\nuesc$ is the fraction
of the particles streaming back from downstream to upstream. 
This quantity is divided by the fraction of particles that would be able
to escape upstream in the absence of waves.
In order not to further increase the complexity
of our numerical simulation, we use here the following approximation
of the representation given in Malkov~(\cite{Malk98}):
\eqb\label{tau}
\tauesc(v,u_2)&=&H\left[ \tilde{v}-(1+\ep) \right]
        \left(1-\frac{u_2}{v}\right)^{-1}\,
        \left(1-\frac{1}{\tilde{v}}\right)\nonumber\\
 & & \cdot\exp\left\{-\left[\tilde{v}-(1+\ep)\right]^{-2}\right\}\,,
\eqe
where the particle velocity is normalized to $\tilde{v}= vk_0/\omega_{\per}$ 
and $H$ is the Heaviside step function.
We argued above that $\omega_{\per}/k_0\simeq u_2/\ep$ 
(see Malkov~\cite{Malk98}, Eq.~42).
The transparency function now solely depends on
the shock velocity in the downstream flow frame, $u_2$, the particle velocity,
$v$, and the relative amplitude of the wave, $\ep$. 

The calculation of the transparency function and the wave--amplitude $\ep$ uses
the ergodicity of the downstream phase--space for the randomized motion of
particles in the high--amplitude wave field. The upstream wave field 
is generated by a beam of leaking particles whose energy density is calculated
from the corresponding area of the downstream phase--space.
From the energy density of this beam the upstream magnetic--field wave 
amplitude is determined self--consistently
(Malkov~\cite{Malk98}).\footnote{Especially downstream, 
where $B_0/B_{\per}\ll 1$, 
the particle--wave interaction can by no means described by 
quasi--linear theory, which would be valid in the opposite case, i.e. for
incoherent waves with small amplitudes. In fact, the calculation of the 
transparency function is {\em not} based on results of the quasi--linear 
theory (Malkov~\cite{Malk98}; Malkov, private communication).
This point is noteworthy, because of existing mis--interpretations of the 
work of Malkov~(\cite{Malk98}) in the literature (Baring~\cite{Bari99}).}
Because of this feedback Malkov was able to constrain the quantity $\ep$
as $0.3\la\ep\la 0.4$, leaving essentially no free parameter.
\begin{figure}[t]
    \vspace{-4cm}
    \begin{center} 
      \epsfxsize9.0cm 
      \mbox{\epsffile{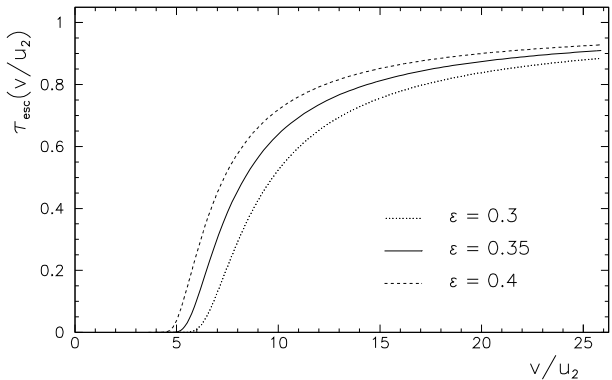}}
    \end{center}
    \vspace{-1cm} 
    \protect\caption{ Transparency function Eq.~(3) vs.~the normalized 
particle velocity $\tilde{v}= v/u_2$ for different values of 
 $\epsilon$. We used the relation $\omega_{\per}/k_0\simeq u_2/\ep$
(see text).}
    \label{tau_plot_2}
\end{figure}
Comparison with hybrid plasma simulations suggests $0.25\la\ep\la 0.35$ 
(Malkov \& V\"olk \cite{MaVo98}), consistent with their analytical results.
This constraint on the wave--amplitude $B_{\per}/B_0 = 1/\ep$ defines
the level to which the particle--wave interaction adjusts.
With the estimation of this amplitude there is no free parameter
describing the level of the beam strength for the injection, and, 
therefore, the injection efficiency. The advantage of our approach presented 
here is that quantities like the plasma velocity and particle momentum 
distribution are calculated
self--consistently by solving simultaneously the hydro--dynamical equations
together with the cosmic--ray transport equation (see Sect.~\ref{dynamics}). 

The function (\ref{tau}) is plotted in Fig.~\ref{tau_plot} 
for $\ep=0.35$ vs.~the particle velocity normalized to 
 $\tilde{v}= vk_0/\omega_{\per}$. In Fig.~\ref{tau_plot} we have also 
reproduced this function as given in Fig.~2 of 
Malkov \& V\"olk~(\protect\cite{MaVo98}) to allow a direct 
comparison. The strong velocity dependence and also the asymptotic behavior 
is modeled reasonably well by the representation Eq.~(\ref{tau}).
In the normalization of Fig.~\ref{tau_plot}, the dependence on $\ep$ is very
weak, and, therefore, not shown.
To illustrate the dependence of the transparency function on small variations 
of the field amplitude, it is better to choose a different normalization.
Therefore, the transparency function Eq.~(\ref{tau}) is shown in 
Fig.~\ref{tau_plot_2} vs.~the velocity normalized to
 $\tilde{v}= v/u_2$ for the maximal allowed range in 
$\ep$, as described above. 

\section{Model}
\label{dyn_model}

\subsection{Dynamical equations}
\label{dynamics}
The standard hydro--dynamical
equations of mass, momentum and energy conservation for a gas with
velocity $u(x,t)$, and density $\rho(x,t)$, corrected for CR pressure
effects are given by
\eqb
\frac{\md \rho}{\md t}&=&-\rho\frac{\partial u}{\partial x}\,,\label{gas1}\\
\rho\frac{\md u}{\md t}&=&
           -\frac{\partial}{\partial x}(P_{\rm g}+P_{\rm c})\,,\label{gas2}\\
\rho\frac{\md e_{\rm g}}{\md t}&=&-\frac{\partial}{\partial x}
            \left[u(P_{\rm g}+P_{\rm c})\right]
             +P_{\rm c}\frac{\partial u}{\partial x}-S(x,t)\,,\label{gas3}
\eqe
where $P_{\rm g}$ and $P_{\rm c}$ are the gas and the CR pressure,
respectively, and 
$e_{\rm g} = {P_{\rm g}}/{\rho}(\gamma_{\rm g}-1)+ u^2/2$ 
is the total energy density of the gas per unit mass.
Here 
${\md}/{\md t}\equiv{\partial}/{\partial t}+
        u{\partial}/{\partial x}\,$
is the total Lagrangian time derivative. 
We assume $\gamma_{\rm g}=5/3$ for the thermal gas adiabatic index throughout
this work. 
The injection energy loss term $S(x,t)$ accounts for the
energy transferred to high energy particles and will be discussed later. 
Equations (\ref{gas1})--(\ref{gas3}) are solved using a
Total Variation Diminishing (TVD) code based on the scheme of
Harten~(\cite{Hart83}).

We assume that the shock Mach number $M=u_{\rm s}/c_{\rm s}$
(with $c_{\rm s}=(\gamma P/\rho)^{1/2}$ the upstream sound speed) exceeds
the Alfv\'en Mach number $M_{\rm A}=u_{\rm s}/\ca\ll M$
(with $\ca=B/(4\pi\rho)^{1/2}$ the upstream Alfv\'en speed).
Then the diffusion--convection
equation, which describes the time evolution of the phase--space density 
$f(p,x,t)$ of the high energy CRs (e.g.~Skilling~\cite{Skil75}),
takes the form:
\eqb\label{diffconv}
\frac{\md f}{\md t}&=&\frac{1}{3}
      \frac{\partial u}{\partial x}
       p\,\frac{\partial f}{\partial p}+ \frac{\partial }{\partial x}
	\left(\kappa(x,p)\frac{\partial}{\partial x}f\right)\,.
\eqe
The diffusion coefficient $\kappa(p,x)$ is assumed to be a scalar. 
Transforming to the variables $y:=\ln(p)$ and $g(y,x,t):=p^4 f(p,x,t)$  
({\it cf.} Falle \& Giddings~\cite{FaGi87}; Kang \& Jones~\cite{KaJo91}),
Eq.~(\ref{diffconv}) can be written as
\eqb\label{CRequation}
\frac{\md g}{\md t}&=&\frac{1}{3}
      \frac{\partial u}{\partial x}\left(
       \frac{\partial g}{\partial y}-4g\right)  
        + \frac{\partial }{\partial x}
	\left(\kappa(y,x)\frac{\partial}{\partial x}g\right)\,.
\eqe
This equation is solved using an implicit Crank--Nicholson scheme,
which is second order in space and time 
(see e.g.~Falle \& Giddings~\cite{FaGi87}).

The high energy particles provide an additional pressure to the system
that has to be included in the set of hydro--dynamical equations
(with $p$ normalized to the proton momentum $p/m_{\rm p}c\to p$): 
\eqb\label{cr_pressure}
P_{\rm c}&=& \frac{4}{3}\pi 
          m_{\rm \scriptscriptstyle p}c^2
          \int\limits_{0}^{\infty}
          (f-\fmax)\,p^4\frac{\md p}{\sqrt{p^2+1}}\,.
\eqe
This definition of the CR pressure $P_{\rm c}$ includes the difference of the 
phase--space density from the Maxwellian distribution $\fmax$ and defines 
the sub--population which we identify as CRs.
The CR energy density is defined as 
\eqb\label{cr_energy}
E_{\rm c}&=& 4\pi m_{\rm \scriptscriptstyle p}c^2
          \int\limits_{0}^{\infty}
          (f-\fmax)\,p^2 (\sqrt{p^2+1}-1)\,\md p\,.
\eqe

\subsection{Injection scheme}
\label{injs}

We do not include an additional injection term in Eq.~(\ref{CRequation}),
because in our model injection is described self--consistently
from the thermal distribution.
Therefore, the lower boundary of the momentum distribution of
the CR population must match the upper boundary of the momentum distribution
of the gas. The distinction between these populations is, of course,
only technical, and defined by the validity of the relevant dynamical
equations. We use a Maxwell
distribution according to the actual density and temperature of the plasma. 
Instead of a fixed
momentum boundary we use here the transparency function $\tauesc$ 
to define where the lower boundary of the CR momentum distribution
matches the momentum distribution of the bulk plasma. The injection
into the high energy part of the phase--space distribution (i.e.~the CRs)
is then directly provided by the bulk of the plasma. 

The initial Maxwellian phase--space density $\fmax(p,x,t)$ is given by:
\eqb\label{maxwell}
\gmax(p,x,t) &=& p^4\fmax(p,x,t)\nonumber\\
 &=& \frac{n(x,t) p^4}{(2\pi m_{\rm p}\,\boltz T)^{3/2}}
                 \exp\left(\frac{-p^2}{2m_{\rm p}\,\boltz T}\right),
\eqe
where $n(x,t)=\rho/m_{\rm p}$ is the particle number density,
and the temperature
is defined by the local gas pressure $P_{\rm g}$ and density $\rho$
according to $T= \mu m_{\rm p} P_{\rm g}/\rho \boltz$.
Here $\mu$ is the mean molecular weight which is assumed to be one, and
$\boltz$ is the Boltzmann constant.
The details of how the momentum distribution is calculated in a time step
from $t-\Delta t$ to $t$ are as follows. 
First we define the CR part of the momentum distribution by 
$\gcr(t-\Delta t)=g(t-\Delta t)-\gmax(t-\Delta t)$.
Now the CR diffusion--convection equation (Eq.~\ref{CRequation}) is 
solved for the entire momentum space, 
including the thermal Maxwell distribution, 
to find the updated distribution function $\tilde{g}(p,x,t)$.
For momenta below the critical momentum of $\tauesc(p_{\rm crit})\equiv 0$ 
any particle acceleration must
be suppressed, and therefore the result of Eq.~(\ref{CRequation}) is rejected
by restoring the Maxwellian distribution given in Eq.~(\ref{maxwell}). 
For momenta above the critical momentum and upstream of the shock,
we use the transparency function as a filter for $\tilde{g}(p,x,t)$ as
described below, since $\tauesc(p,t)$ corresponds to 
the fraction of the phase--space density at a given momentum that can cross 
the shock from downstream to upstream. 
The final distribution at time $t$ is then
given by $g(p,x,t)$ in the following way:
\eqb
\gcrt(p,x,t) &=& \tilde{g}(p,x,t)-\gmax(p,x,t)\,,\\
\gcr(p,x,t) &=& \gcr(p,x,t-\Delta t)+ \tau_{\rm esc}(p,t)\nonumber\\
 & & \cdot\left[\gcrt(p,x,t)
                         -\gcr(p,x,t-\Delta t)\right]\,,\\
g(p,x,t) &=& \gcr(p,x,t)+\gmax(p,x,t)\,.
\eqe
So effectively at the lower momentum limit where $\tauesc(p)\equiv 0$ 
the Eq.~(\ref{CRequation}) has no effect at all,
while at the higher momentum limit where $\tauesc(p)\equiv 1$
the result of Eq.~(\ref{CRequation}) is used without
further modification.
Only in the intermediate momentum regime where $ 0< \tauesc(p) < 1$, 
the transparency function represents the injection process (i.e. thermal leakage).
Thus the transparency function
defines self--consistently the momentum boundary above which the
particle acceleration mechanism can work, and also defines the transition region
between thermal plasma and accelerated particles.

The particle injection rate into the CR population can
be estimated from the adiabatic change of 
the momentum due to the velocity gradient of the flow:
\eqb\label{Q}
Q(p,x,t) &=& 4\pi\, p^2 f(p,x,t)\,\left(\frac{\partial p'}
 {\partial t}\right)_p\nonumber\\
 &=& -\frac{4\pi}{3}p^3f(p,x,t)\frac{\partial u}{\partial x}\,.
\eqe
Then the energy loss rate of the gas can be written as
\eqb\label{S}
S(x,t)&=&
\frac{1}{2} m_{\rm p} c^2 \int\limits_{0}^{\infty}
     \frac{\partial \tauesc(p,t)}{\partial p} p^2 Q(p,x,t) \,\md p\,,\\ 
&=& -\frac{2}{3}\pi\, m_{\rm p} c^2 \,
    \frac{\partial u}{\partial x}\int\limits_{0}^{\infty}
     \frac{\partial \tauesc(p,t)}{\partial p} p^5 f(p,x,t) \,\md p\,.\nonumber
\eqe
Note here the condition ${\partial \tauesc(p,t)}/{\partial p} \not= 0 $ in fact
defines the ``injection pool'' where the thermal leakage takes place. 
Due to the steep dependence of both the Maxwell distribution 
and the transparency function on the particle momentum, 
the momentum range of the injection pool is well restricted. 
Either below or above this momentum range
$\tauesc(p)=$ constant, so ${\partial \tauesc(p,t)}/{\partial p} = 0 $. 
If the transparency function is given by a step function
$\tauesc(p)=H(p-\pinj)$, it becomes the injection scheme adopted by  
Kang \& Jones (\cite{KaJo95}, \cite{KaJo97}) in which the injection takes 
place at a single injection momentum rather than an extended momentum range. 

The transparency function $\tauesc$ given by Eq.~(\ref{tau}) depends 
on the downstream plasma velocity, which is averaged over the 
diffusion length of the 
particles with momentum at the injection threshold. This dependence is also
important for the injection efficiency, and leads to a regulation mechanism
similar to the above beam wave interaction. If the initial injection is
so strong that a significant amount of energy is transferred from the gas
to high energy particles, the downstream plasma cools, and, in addition, 
the downstream bulk velocity decreases in the shock frame due to the shock 
modification of the cosmic--ray population.
Because the injection pool is in the high energy tail of the Maxwellian
distribution of the gas, the cooling decreases significantly the injection
rate. However, the deceleration, in turn, allows for a modest increase of the 
phase--space of particles that can be injected. This is expressed by the
$u_2$ dependence of Eq.~(\ref{tau}). 
This velocity dependence balances partly the reduction of injection due to 
the cooling of the plasma. Remarkably, these two effects 
lead to a very {\em weak} dependence of the injection efficiency on $\ep$ 
in the vicinity of $\ep\approx 0.35$.
\subsection{Diffusion model}
Since the injection process is included self--consistently,
the diffusion coefficient is the only remaining free parameter in our
model.
We assume the particle diffusion is based on the scattering off the 
self--generated waves 
which have a field component perpendicular to the plasma flow.
The compression of the plasma leads to an amplification of these waves,
which is described by scaling the diffusion coefficient
as $\kappa\propto 1/\rho$. In our one dimensional
model we have to describe diffusion along the mean magnetic field. The
lower limit for the diffusion coefficient is the Bohm
diffusion coefficient
$\kappaB (p)=(3\cdot10^{22}~{{\rm cm}^2}/{\rm s}/ B_{\mu {\rm G}})
        ~{p^2}/{(1+p^2)^{1/2}}\,,$
where $B_{\mu {\rm G}}$ is the magnetic field strength in units of
micro--gauss.
For the present calculations we assume the diffusion coefficient is 
simply related to Bohm diffusion as 
\eqb
 \kappa(p) =\zeta\,~\kappaB\,\rho_0/\rho(x)\,.
\eqe
We have introduced the factor $\zeta$ to
account for the higher diffusion in the direction of the
mean magnetic field, because this direction is parallel to the shock normal,
and, therefore, relevant for the acceleration process at quasi--parallel
shocks.  Although the time scale for the cosmic--ray acceleration
does depend on the diffusion coefficient, the basic self regulation
process for the injection problem which we investigate here is not
dependent on the choice of $\zeta$. Therefore, since our study intends to 
focus on the general time dependent behavior of this injection model,
we do not include a completely
self--consistent scattering model, where the diffusion coefficient is
coupled to the spectrum of the Alfv\'en waves.

\subsection{Initial and boundary conditions}
We assume there is no pre--existing CR population, 
so the initial particle distribution is purely Maxwellian with 
the local plasma temperature and density. 
We use open boundary conditions for the description of the thermal plasma 
in our simulations.
In momentum space, the lower boundary is provided by the Maxwell distribution
as discussed above.
At the highest momentum and also at the upstream boundary in 
configuration--space we use a \lq free escape\rq\ boundary.
Downstream we use a \lq no diffusive flux\rq\ boundary, where
the cosmic--ray density is always kept constant across the boundary.
However, the grid size was chosen so large in the simulations we present
here, that the cosmic--ray pressure is at all times essentially zero at both 
boundaries.

\section{Results for strong shocks} 
\label{results}
First we consider a strong shock with an initial Mach number of $M=30$. 
Unlike an ordinary hydrodynamic simulation, the simulation of the CR 
shock acceleration requires
specification of three physical parameters, $u_0/c$, $\rho_0/m_{\rm p}$, 
and the shock Mach number in addition to the diffusion coefficient.  
We adopted the following nominal physical scales for physical parameters: 
$u_0=5000\,{\rm km\,s}^{-1}$,
$\rho_0/m_{\rm p} = 0.03\,{\rm cm}^{-3}$, 
$t_0=4.0\cdot 10^4$ s,
$x_0=2.0\cdot 10^{13}$ cm, 
$P_{\rm g 0}=\rho_0 u_0^2= 1.25\cdot 10^{-8}\,{\rm erg\,cm}^{-3}$. 
We use $\zeta=100$ for the simulations presented here, and a magnetic
field of $B=3\mu$G.
The initial conditions are specified as follows:
 $\rho_{\rm up}=\rho_0$, 
$u_{\rm up}=- u_0$, and $P_{\rm g,up}= 6.667\cdot 10^{-4} P_{\rm g0}$ in
the upstream region, while $\rho_{\rm d}=3.987\rho_0$, 
$u_2=- 0.25 u_0$, $P_{\rm g,d}= 0.75 P_{\rm g0}$ 
downstream. 
These values reflect
the shock jump conditions in the rest--frame of the shock.

We define the diffusion length and time at a given momentum as
$l_{\rm d}(p) = \kappa(p) / u_0$ and
$t_{\rm d}(p) = \kappa(p) / u_0^2$.
For a proper convergence, the spatial grid size should be smaller than
the diffusion length of the injection pool particles
($l_{\rm d}/x_0 \sim 0.04$ in upstream for $\pinj\sim 0.02$).  
On the other hand, the spatial region of the calculation in upstream 
and downstream should be larger 
than the diffusion length--scale of the particles with the highest energies 
reached at the end of our simulation period
($l_{\rm d}/x_0 \sim 230$ for $p_{\rm max}\sim 2.5$).  
So we used 51200 uniform grid zones for $x/x_0=[-250,250]$, with the shock 
initially at $x=0$ and the grid size $\Delta x/x_0 = 0.01 = 0.25\cdot l_{\rm d}
(\pinj)/x_0$. 
We use 128 uniform grid zones in $\log(p)$ for $\log(p)=[-3.0,0.477]$.
We integrate the solutions until $t=240 t_0$ which corresponds to $t_{\rm d}$  
for $p_{\rm max}\sim 2.5$, so the CR particles
became only mildly relativistic by the end of our simulations. 

\begin{figure}[t]
    \vspace{-0.8cm}
    \begin{center} 
      \epsfxsize9.0cm 
      \mbox{\epsffile{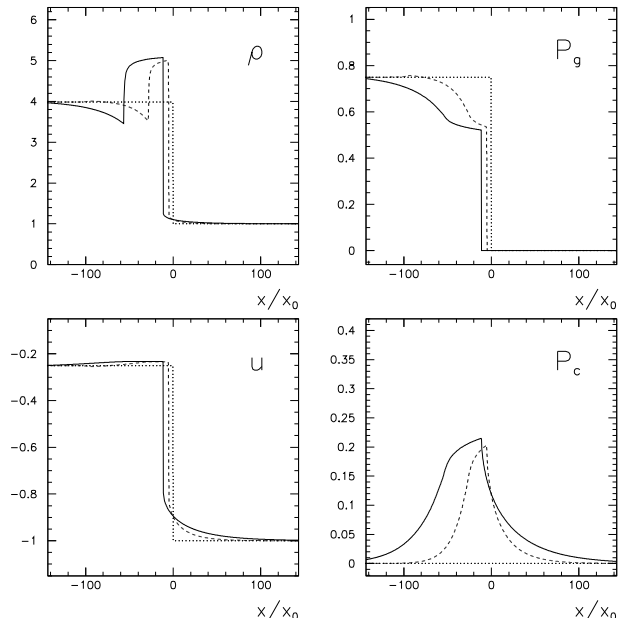}}
    \end{center}
    \vspace{-1.cm} 
    \protect\caption{Gas density $\rho/\rho_0$, pressure 
$P_{\rm g}/P_{\rm g0}$, velocity $u/u_0$, and cosmic--ray pressure 
$P_{\rm c}/P_{\rm g0}$, at times $t=0$ (dotted), $t=120\,t_0$ (dashed) and 
 $t=240\,t_0$ (solid line). The shock Mach number is $M=30.0$, 
 $\epsilon = 0.35$ and $\zeta=100$. The initial upstream gas pressure is
 $P=6.667\cdot 10^{-4}P_{\rm g 0}$.}
    \label{Hi_Mach_plasma}
\end{figure}

\subsection{Dynamical evolution} 
\label{spatial}
Figure \ref{Hi_Mach_plasma} shows the normalized gas density $\rho(x)$, 
gas pressure $P_{\rm g}(x)$,
plasma velocity $u(x)$ and the cosmic--ray pressure $P_{\rm c}(x)$ over 
the spatial length $x$, for different times. 
This shows clearly the basic features of the shock 
modification by a diffusive component; that is, the adiabatic precursor 
compression and the sub--shock. 
The CR pressure $P_{\rm c}$ is 
responsible for the deceleration and compression of the plasma flow 
in the precursor region upstream of the sub--shock, which still remains  
strong. As a result, the gas is compressed to higher density downstream 
of the sub--shock. 

The cosmic--ray pressure immediately downstream of the 
sub--shock has not reached a steady state yet. The reason is that for a the 
non--thermal particles with a momentum distribution $f\propto p^{-s}$ with
$s\le 4$, the energy density is an increasing function of $p_{\rm max}$.
This applies even if the injection is shut down completely, like for 
an $\delta$--function type injection in time, as shown by Drury~(\cite{Drur83}).
We expect that $P_{\rm c}$ will continue to increase after our integration
time $t=240 t_0$, which leads to a significant modification 
of the shock structure and to the steepening of the power--law distribution 
of suprathermal particles.
The simulations of such non--linear evolution, however,
require much greater spatial region and grid zones and also longer 
integration time than what we could afford in our simulations. 

In real astrophysical shocks, the energy density is limited by 
radiation losses in the case of electrons
or more generally by particle escape due to the 
finite extent of the acceleration region. 
For the maximum energy of particles ($ p_{\rm max} \sim 2.5$) 
achieved by $t=240\, t_0$ in this simulation, 
neither effect is important, and, therefore, not 
included.  
\subsection{Energy distribution}
\label{energy}

\begin{figure}[t]
    \vspace{-1.8cm}
    \begin{center} 
      \epsfxsize8.5cm 
      \mbox{\epsffile{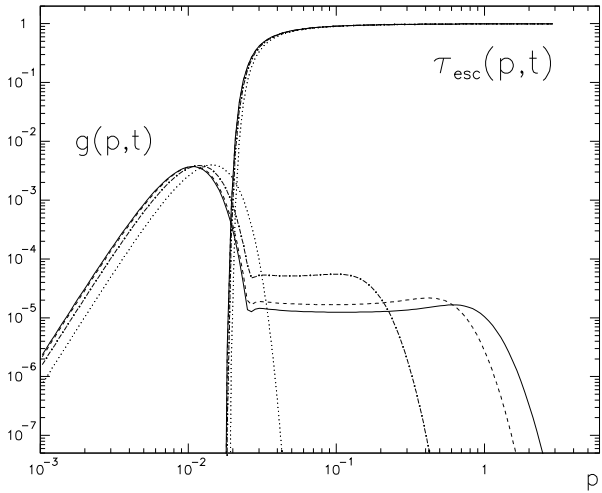}}
    \end{center}
    \vspace{-1cm} 
    \protect\caption{Phase--space density $g=p^4f$ 
 vs.~proton momentum immediately downstream of the sub--shock. 
Also shown is the transparency function $\tauesc$. 
Both functions are presented for $t=0$ (dotted), $t=10\,t_0$ (dot--dashed), 
$t=120\,t_0$ (dashed), and $t=240\,t_0$ (solid line). 
For the parameters used see Fig.~4 and corresponding text.}
    \label{Hi_Mach_g_tau}
\end{figure}

The phase--space distribution $g(p,x,t)\equiv p^4\, f(p,x,t)$ 
immediately (three zones) behind the sub--shock 
is shown in Fig.~\ref{Hi_Mach_g_tau} for three different times. 
Initially this distribution is given by a 
Maxwell distribution, as shown by the dotted line. 
At the thermal part of the distribution the cooling of the postshock 
gas due to the energy flux into the CR particles
is responsible for the shift of the Maxwellian distribution 
towards lower energies.
We have also plotted the transparency function $\tauesc$ at the same 
simulation times. 
According to Eq.~(\ref{S}) the injection rate into the
non--thermal distribution depends on overlap of 
${\partial \tauesc}/{\partial p}$ and $g(p)$ that determines the 
injection pool. 
One can see that initially the injection rate is high and so
the postshock gas cools quickly, resulting in narrowing down of 
the injection pool. This causes the injection rate to decrease.
But then the transparency function also shifts toward lower momenta,
because the downstream plasma velocity $u_2$ decreases as 
the postshock gas cools.
The combination of the shift of $\tauesc$ toward lower momenta 
and the decrease of the particles in the Maxwellian tail due to 
the gas cooling leads to the self--regulation of the injection rate 
at a quite stable value. 
According to the plot of $g(p)$ at $t=120\,t_0$ and $t=240\,t_0$,
the Maxwell distribution turns into a power--law at an almost constant  
``effective injection momentum'' which determines the magnitude
of the CR distribution function $g(p)$ at a stable value 
(about $1/200$ of the thermal peak).
The value of this constant effective injection momentum can be
translated into the parameter $c_1= \pinj/ \pth \sim 2.3$ 
(where $\pth = 2\sqrt{ m_{\rm p} \boltz T}$) 
defined by Kang \& Jones (\cite{KaJo95}).  
But this is somewhat larger 
than what they used ($c_1=1.9-1.95$). 

\begin{figure}[t]
    \vspace{-1.6cm}
    \begin{center} 
      \epsfxsize10.0cm 
      \mbox{\epsffile{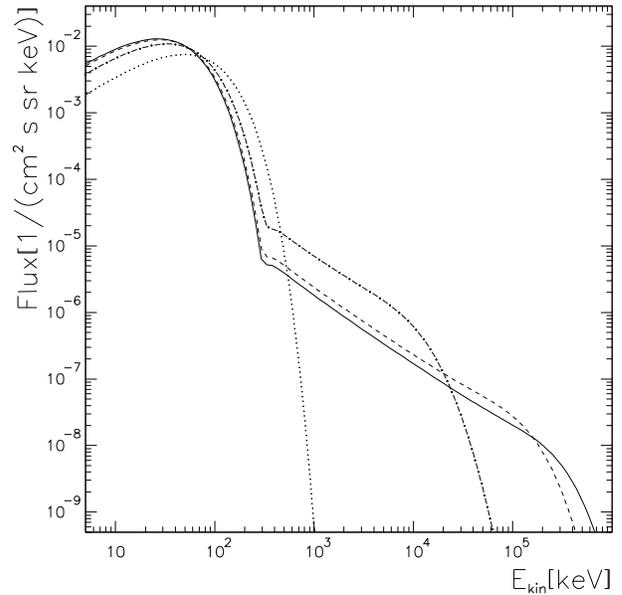}}
    \end{center}
    \vspace{-1.4cm} 
    \protect\caption{Omni--directional flux vs.~proton kinetic
energy, for $t=0$ (dotted), $t=10\,t_0$ (dot--dashed), $t=120\,t_0$ (dashed), 
and $t=240\,t_0$ (solid line). These distributions are identical to
those shown in Fig.~5. For the parameters used see 
Fig.~4 and corresponding text.}
    \label{Hi_Mach_flux}
\end{figure}

The narrow injection pool also leads to 
a rather sharp transition from the Maxwell distribution to 
the non--thermal part
starting shortly above the effective injection momentum
(see Fig.~\ref{Hi_Mach_g_tau}). 
The canonical result in the test particle limit, 
$g(p)=p^4\, f(p)=p^4\, p^{-s}=$ constant, 
for a strong shock with $s=3r/(r-1)=4$ is reproduced very well
in our simulations. 
The same energy spectrum is shown in Fig.~\ref{Hi_Mach_flux}
in the form of the  omni--directional flux
 $F(E)\md E \propto v\,p^2 f(p)\md p$ vs.~proton kinetic 
energy downstream of the shock normalized to $t=0$. 
At energies above the injection pool we expect, for the strong shock ($r\simeq 4$)
simulated here, the result $F(E)\propto E^{-\sigma}$, with
 $\sigma=\{(r+2)/(r-1)\}/2=1$, which 
is reproduced with high accuracy. 

In using the standard cosmic--ray transport equation, we have, of course, made
use of the diffusion approximation, which may introduce an error especially for
$v\simeq u_2$. Using an eigenfunction method, 
Kirk \& Schneider (\cite{KiSc89}) have explicitly
calculated the angular distribution of accelerated particles and accounted for
effects of a strong anisotropy especially at low particle velocities.
They were able to calculate the injection efficiency without recourse to
the diffusion approximation, and found always lower efficiencies compared
to those in the diffusion approximation.
Using the initial thermal distribution, 
we have estimated an effective injection momentum from the peak of
the distribution function, $\gmax(p) \tauesc(p)$.
For the shock parameters considered here and for
$\ep=0.35$ we get an effective initial injection velocity of 
about $6700\,{\rm km\, s}^{-1}$ (in the shock frame). 
For this injection velocity, $r=4$ and $u_0=5000\,{\rm km\, s}^{-1}$, 
they estimate
a reduction effect of $\approx 8\%$, leaving the diffusion approximation 
as quite reasonable even in this regime.
\subsection{Injection and acceleration efficiencies}
\label{efficiency}
To describe the injection efficiency often a parameter $\xi$ is 
used for the fraction of the in--flowing plasma particles that 
are instantaneously
accelerated to a fixed injection momentum $\pinj$ 
(e.g.~Falle \& Giddings~\cite{FaGi87}; Dorfi~\cite{Dorf90}; 
Jones \& Kang~\cite{JoKa90}; 
Zank et al.~\cite{ZaWeDo93}; Berezhko et al.~\cite{BeYeKs94}). 
The injection energy flux $I$ transferred to CRs is then given by
\eqb\label{I}
I=\xi\,\frac{\rho_1 u_1}{m}\,\frac{\pinj^2}{2m} \,,
\eqe 
where $u_1$ is the upstream plasma velocity in the shock frame, and 
$\rho_1$ is the upstream density.
From the fact that the injected energy flux $I$ must be equal to 
the spatial integral of the injection energy loss term $S(x)$,
that is, $ I = \int S(x)\,\md x$ and  
by assuming momentarily a step function 
for the transparency $\tauesc(p)=H(p-\pinj)$, 
we get:
\eqb
\xi(t)=-\frac{4 \pi}{3}\,\int\frac{\partial u}{\partial x}
        \frac{\pinj^3 f(\pinj)}{n_1\,u_1}\,\md x\,,
\eqe
where $n_1=\rho_1/m$ is the upstream number density. This is equivalent to
the injection parameter used by Kang \& Jones~(\cite{KaJo95}).
The so--defined injection parameter $\xi$ is, however, not an exact 
measure of the number of particles contributing to the population of cosmic
rays, because the acceleration process cannot be described by shifting
particles instantaneous from thermal energies to an injection momentum $\pinj$.
Furthermore, $\xi$ depends strongly on the chosen injection momentum $\pinj$, 
which is not a fixed single parameter in our numerical simulation.

\begin{figure}[t]
    \vspace{-1cm}
    \begin{center} 
      \epsfxsize10cm 
      \mbox{\epsffile{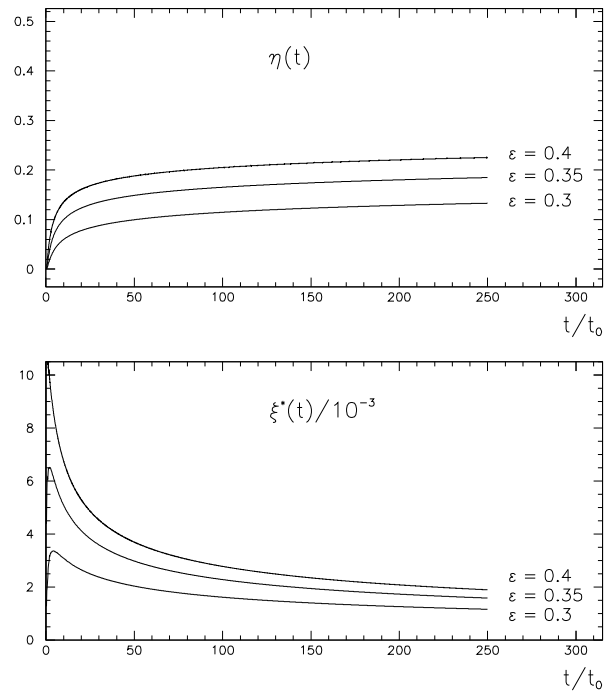}}
    \end{center}
    \vspace{-0.8cm} 
    \protect\caption{ Energy efficiency $\eta(t)$ and
the fraction of cosmic--ray particles $\xi^*(t)$ for three values of
the inverse wave--amplitude $\epsilon$ at a strong, $M=30$, shock.
For the parameters used see Fig.~4 and corresponding text.}
    \label{Hi_Mach_eff}
\end{figure}

A method to measure the injection efficiency without specifying the injection
momentum, is to compare the number of particles in the CR part 
to the number of particles swept through the shock.
According to our definition of the CR population we can write
for the CR number density
\eqb
\ncr(x,t) &=& \int\fcr(p,x,t)\md^3 p \nonumber\\
 & \equiv &  \int\left(f(p,x,t)-\fmax(p,x,t)\right)\md^3 p\,.
\eqe
The fraction of particles that has been swept through the shock after the 
time $t$, and then injected into the cosmic--ray distribution is then 
given by
\eqb
\xi^*(t)=\frac{\int\ncr(x,t)\md x}{n_1 u_1\,t}\,.
\eqe
The time development of this injection efficiency is plotted 
in Fig.~\ref{Hi_Mach_eff} for three values of the inverse wave 
amplitude $\epsilon$. Recall that Malkov~(\cite{Malk98}) found
 $0.3\la\ep\la 0.4$. 
In the very beginning
of the simulation the injection does depend strongly on the wave--amplitude,
because of the very steep dependence of the Maxwell distribution 
at the injection energies. However, as described above, a strong initial 
injection leads to a temperature decrease of the plasma, and to a shift 
of the Maxwell distribution, which balances this effect. Therefore at later
times the fraction of injected particles, $\xi^*$, does not depend strongly
on the initial wave--amplitude. At time $t/t_0 = 250$ (or $t=1.0\cdot 10^7$s)
we get a fraction of injected particles of $\xi^*=(1.5\pm 0.4)\cdot 10^{-3}$ 
for the interval $\ep=0.35\pm 0.5$. 

To measure the efficiency of the particle acceleration at a shock front,
we compare the energy flux in cosmic rays to the total energy which is 
available from the downstream plasma flow. This energy consists of
the sum of kinetic energy and the gas enthalpy. The fraction of this initial 
energy flux, which is transferred to CRs is given by 
\eqb\label{eta}
\eta(t)&=&\frac{\frac{\gac(t)}{\gac(t)-1}u_{\rm d}(t) P_{\rm c}(t)}
         {\frac{1}{2}\rho_{\rm d} u_{\rm d}^3 + \frac{\gag}{\gag-1} u_{\rm d} 
          P_{\rm g,d}}\,,
\eqe
where $u_{\rm d}=u_{\rm s}(1-1/r)$ is the initial downstream plasma velocity 
in the upstream rest frame.
The definition of the efficiency $\eta(t)$ is similar to the definition from
V\"olk et al.~(\cite{VoDrMc84}). However, Eq.~(\ref{eta}) compares
the energy flux in CRs not only to the kinetic energy flux 
of the gas, but also includes the gas enthalpy flux.
We measure the CR pressure immediately downstream of the sub--shock,
where it will first reach the constant downstream value, in case a steady 
state does exist (see below).
The time dependent values  are averaged over the interval
$x/x_0 = [-0.5 u_2 t,\dots ,0]$ in the shock frame to avoid influence 
of small scale modifications
of the cosmic--ray pressure and plasma velocity on the injection efficiency. 
When the quantities $u_2$, $P_{\rm c}$ and $\gamma_{\rm c}$ have 
reached steady--state distributions downstream of the sub--shock, $\eta(t)$ 
is also no longer time dependent.

The evolution of the energy efficiency, $\eta(t)$, 
is plotted in Fig.~\ref{Hi_Mach_eff} for three
different magnetic--field wave amplitudes. 
See Fig.~\ref{Hi_Mach_plasma} and the 
description in Sect.~\ref{results} for the corresponding parameters.
The case $\epsilon=0.4$ corresponds to the highest injection efficiency
and therefore leads to the highest cosmic--ray pressure.
To assure a vanishing value of the cosmic--ray pressure at the spatial grid 
boundaries at all times, the calculation for $\epsilon=0.4$ was done on a 
somewhat larger grid with 60416 uniform zones for $x/x_0=[-300,300]$.
For the value $\epsilon=0.35$, which 
was calculated by Malkov~(\cite{Malk98}), we see that about 20\% of the 
available energy in this shock is transferred into the cosmic--ray population.
The acceleration efficiency has, however, not reached a real steady state
value, but is increasing with $\eta(t)\propto t^\alpha$ 
with $\alpha\approx 0.1$. 
The acceleration efficiency achieved by this time
is given by $\eta=(18\pm 5)\%$ for $\ep=0.35\pm 0.5$. 
Thus a substantial amount of the initial energy flux at a
shock front can be transferred to a high energy part of the distribution,
during the relatively short time we have simulated here.

\section{Results for weak shocks}
When the initial compression ratio decreases for a  weak shock, 
the injection process is influenced
in several ways by the change in the plasma and magnetic field properties.
To investigate the effects of a lower compression ratio and lower Mach
number on the injection process we will consider an example with $r=2.5$
and $M=2.24$.
At such a shock, the phase space for which the downstream particles can 
re--cross the shock to upstream is decreased compared to the strong shock 
case, because the shock velocity in the downstream rest frame
 $u_2=u_{\rm s}/r$ is inversely proportional to the compression ratio.
At the same time the plasma is heated less, because
the transformation of kinetic energy to thermal energy depends also on 
the compression ratio;
 $\Delta k_{\rm \scriptscriptstyle B} T\propto  
 m_{\rm \scriptscriptstyle P}u_{\rm s}^2(1-1/r^2)$. This shifts the
downstream Maxwell distribution to lower energies, as compared to higher 
compression, and, therefore, influences strongly the number of particles in 
the momentum range making the potential injection pool.
\label{resultsweak}
\begin{figure}[t]
    \vspace{-0.8cm}
    \begin{center} 
      \epsfxsize9.0cm 
      \mbox{\epsffile{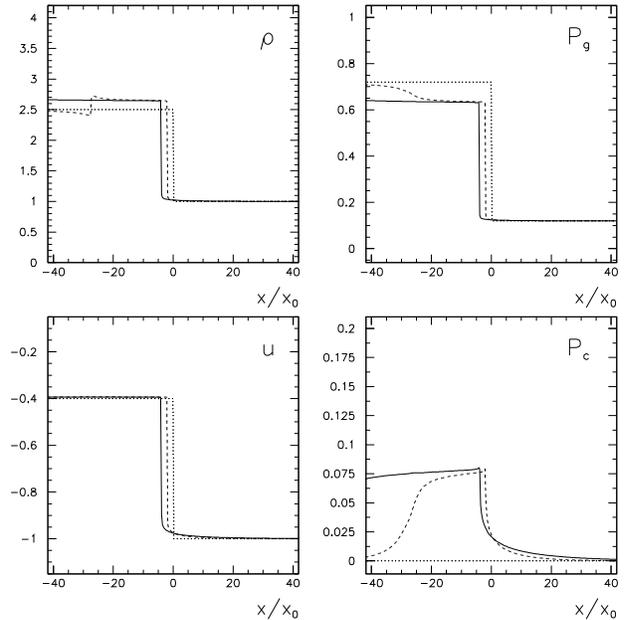}}
    \end{center}
    \vspace{-1cm} 
    \protect\caption{Gas density $\rho/\rho_0$, pressure 
$P_{\rm g}/P_{\rm g0}$,
velocity $u/u_0$, and cosmic--ray pressure $P_{\rm c}/P_{\rm g0}$, 
at times $t=0$ (dotted), $t=70\,t_0$ (dashed) and 
 $t=140\,t_0$ (solid line). The shock Mach number is $M=2.24$, 
 $\epsilon = 0.6$, and $\zeta=100$. The initial upstream gas pressure is
 $P=0.12 P_{\rm g 0}$.}
    \label{Lo_Mach_plasma}
\end{figure} 
On the other hand, at quasi--parallel shocks, the amplitude
of the magnetic field wave spectrum $B_{\per}$ is amplified downstream
by the factor $r$.  For a decreasing compression ratio, the downstream
plasma becomes more transparent. This balances the effects
of the phase space and temperature changes described above.
The initial downstream (inverse) wave--amplitude
 $\ep = B_0/B_{\per}$ was calculated to be in the interval 
$\ep\approx[0.3,\dots , 0.4]$ in the limit of strong shocks
(Malkov~\cite{Malk98}; Malkov \& V\"olk~\cite{MaVo98}).
An extrapolation to weak shocks with $r=2.5$ of this interval by multiplying
$\ep$ with the factor of (4/2.5)  gives $\ep\approx[0.48, \dots , 0.64]$. 
However, the 
calculation of the transparency function was based on the assumption of an 
high amplitude wave spectrum downstream ($\ep \ll 1$). With decreasing
wave amplitude the velocity dependence of the transparency function 
changes towards its asymptotic function, defined by particle kinematics
without a wave field:
$\tau(v)=0$ for $v < u_2$ and $\tau(v)=1$ for $v\ge u_2$.
On the other hand, this limit may be reached in reality only if the
resulting beam from downstream to upstream is too weak to produce
a magnetic field instability.

As an initial exploration of this behavior, we will present here results for 
the spatial and momentum distributions
and the energy and particle injection efficiency for an inverse magnetic fields
amplitude parameter $\ep$ in the range $\ep=[0.4, \dots , 0.7]$.
We have included the value $\ep = 0.4$ to compare the results directly to
the strong shock case. This can demonstrate the principal effect of weaker
shocks on the injection process.
The resulting injection efficiencies and shock modifications for all values
of $\ep$ shown here should be
considered as lower limits for the weak shock with $r=2.5$ ($M=2.24$) 
as described above. 

\begin{figure}[t]
    \vspace{-1.8cm}
    \begin{center} 
      \epsfxsize8.5cm 
      \mbox{\epsffile{gieseler.f9}}
    \end{center}
    \vspace{-1cm} 
    \protect\caption{Phase--space density $g=p^4f$ vs.~proton momentum 
immediately downstream of the sub--shock. Also shown is the transparency 
function $\tauesc$. Both functions are presented for $t=0$ (dotted),
$t=10\,t_0$ (dot--dashed), $t=70\,t_0$ (dashed), and $t=140\,t_0$ (solid line).
For the parameters used see 
Fig.~8 and corresponding text.}
    \label{Lo_Mach_g_tau}
\end{figure}

The physical scales are specified as follows: $t_0=1.11\cdot 10^5$ s,
$x_0=3.33\cdot 10^{13}$ cm, $u_0=3000\,{\rm km\,s}^{-1}$,
 $\rho_0/m_{\rm p} = 0.03\,{\rm cm}^{-3}$, 
$P_{\rm g 0}=4.52\cdot 10^{-9}\,{\rm erg\,cm}^{-3}$. 
We use $\zeta=100$ for the simulations presented here, and a magnetic
field of $B=3\mu$G.
The initial values for the $M=2.24$ case are $\rho_{\rm up}=\rho_0$,
$u_{\rm up}=- u_0$,  and $P_{\rm g,up}= 0.12 P_{\rm g0}$ 
in the upstream region,
while $\rho_{\rm d}=2.5\rho_0$,
$u_2=- 0.4 u_0$, and $P_{\rm g,d}= 0.72 P_{\rm g0}$
in the downstream.
We have used 44032 uniform grid zones for $x/x_0=[-170,130]$, 
with the shock 
initially at $x=0$, and 128 uniform grid zones in $\log(p)$ 
for $\log(p)=[-3.0,0]$.
The corresponding Mach number is $M=2.24$. 

Figure~\ref{Lo_Mach_plasma}
shows the normalized gas density $\rho(x)$, gas pressure $P_{\rm g}(x)$,
plasma velocity $u(x)$ and the cosmic--ray pressure $P_{\rm c}(x)$ over 
the spatial length $x$, for different times. 
Because the resulting non--thermal spectrum produced as a result
of the injection and particle acceleration is steeper than in the strong shock
case, the pressure $P_{\rm c}$ 
in this distribution remains small compared to the gas pressure at all times. 
As a result, the shock is
modified only slightly. Also the temperature of the downstream plasma remains
almost constant. Furthermore, because the energy density in non--thermal
particles is not an increasing function in time, the shock modification 
can reach a steady state earlier, as compared to the strong shock case. 
In fact, at time $t=140 t_0$, 
shown in Fig.~\ref{Lo_Mach_plasma}, the pressure $P_{\rm g}$, $P_{\rm c}$,
the velocity $u$ and the density $\rho$ immediately downstream has
reached almost a steady state. 

\begin{figure}[t]
    \vspace{-1cm}
    \begin{center} 
      \epsfxsize10cm 
      \mbox{\epsffile{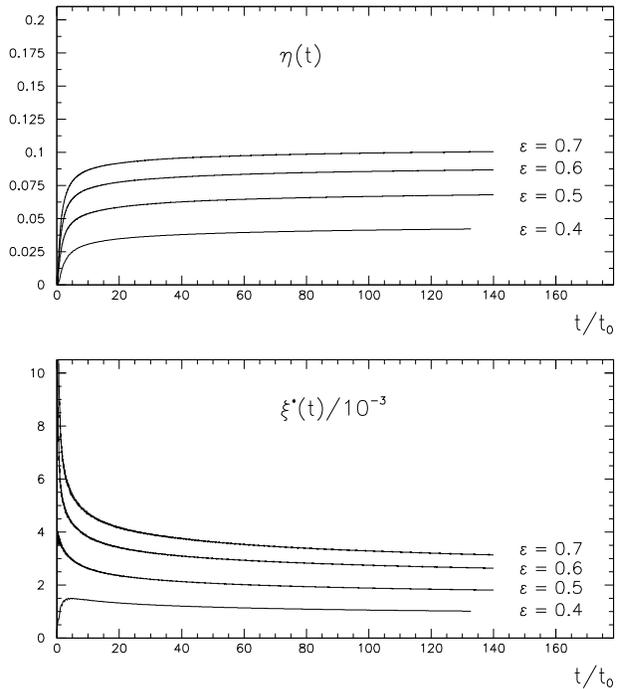}}
    \end{center}
    \vspace{-0.8cm} 
    \protect\caption{Energy efficiency $\eta(t)$ and
the fraction of cosmic--ray particles $\xi^*(t)$ for four values of
the inverse wave--amplitude $\epsilon$ at a weak shock.
For the parameters used see Fig.~8 and corresponding text.}
    \label{Lo_Ma_eff}
\end{figure}

The downstream momentum distribution in Fig.~\ref{Lo_Mach_g_tau} shows
clearly the steeper spectrum of the non--thermal part, which asymptotes to 
the standard result $g(p)\propto p^{-s+4}$ with $s=3r/(r-1)=5$ for $r=2.5$.
It can be seen also, that the thermal part of the distribution is not as 
much modified as in the strong shock case (compare Fig.~\ref{Hi_Mach_g_tau}).
Because the modification of the transparency function over time depends
only on changes in the downstream plasma velocity, it remains essentially
unchanged.

The energy efficiency $\eta(t)$, as defined in Eq.~(\ref{eta}), is lower 
roughly by a factor of two compared to the strong shock case, 
because of the steeper non--thermal spectrum and the resulting energy density 
(compare Fig.~\ref{Hi_Mach_eff} and Fig.~\ref{Lo_Ma_eff}).
Our results for the wave amplitude, $0.5\le\ep\le 0.7$,
give the injection efficiency, $\xi^*= (2.5\pm 0.7)\cdot 10^{-3}$
at time $t=140 t_0 = 1.55\cdot 10^7$s, where the time evolution can be
considered as almost a steady state. 
The number of particles, which are in the non--thermal part is comparable 
to the strong shock considered above at this time. 
In addition, we point out that the application
of the above described injection model to weak shocks is an extrapolation,
and we believe would yield lower limits on the injection efficiency. 
\section{Conclusions}
We have developed a numerical method to include self--consistently the
injection of the supra--thermal particles into the cosmic--ray population
at quasi--parallel shocks according to the analytic solution of 
Malkov (\cite{Malk98}).
Toward this end, we have adopted the ``transparency function'' 
$\tauesc(v,u_2)$ which expresses the probability that supra--thermal 
particles at a given velocity can leak upstream through the magnetic waves, 
based on non--linear particle interactions with self--generated waves. 
We have incorporated the transparency function
into the existing numerical code which solves the
cosmic--ray transport equation along with the gas dynamics equations. 
In order to investigate the interaction of high energy particles, 
accelerated by the Fermi process, with the underlying plasma flow 
{\em without} using a free parameter for the injection efficiency,
we have applied our code with the new injection scheme to both strong 
($M=30$) and weak ($M=2.24$) parallel shocks.  

The main conclusions from the simulation results are as follows: 

\begin{enumerate}

\item{}
The injection process is regulated by the overlap of 
the population of supra--thermal particles in the injection pool
and the function ${\partial \tauesc(p,t)}/{\partial p}$. 
As being in the high energy tail of the Maxwell velocity distribution,
the population in the injection pool depends
strongly on the gas temperature and the particle momentum.  
The function ${\partial \tauesc(p,t)}/{\partial p}$ behaves like a
delta--function defined near a narrow injection pool.  
As the postshock gas cools due to high initial injection, the Maxwell
distribution shifts to lower momenta.
But the transparency function also shifts to lower momenta, as well,
due to its dependence on the postshock flow velocity.
As a result, the injection rate reaches and stays at a stable value 
after a quick initial adjustment, and also depends only weakly on the initial 
conditions. 
This self--regulated injection may imply a broad application of our
simulation methods. 

\item{}
The fraction of the background particles that are accelerated to form the 
non--thermal part of the distribution turns out to be in the range
 $1.2\cdot 10^{-3} \la \xi^* \la 1.9\cdot 10^{-3}$ 
for the range of initial wave--amplitudes 
$0.3\le\ep\le 0.4$ at a $M=30$ shock. 
For a $M=2.24$ shock, a slightly higher injection is achieved
at $\xi^*= (2.5\pm 0.7)\cdot 10^{-3}$, but this could be a lower limit.
Such values for the particle injection efficiency have been used as a 
parameter for spherically expanding SNRs 
by several authors (Dorfi~\cite{Dorf90}; Jones \& Kang~\cite{JoKa92};
Berezhko et al.~\cite{BeKsYe95}; Berezhko \& V\"olk~\cite{BeVo00}). 
These values are well above the ``critical injection rate'' of 
$\eta_{\rm crit} \sim 10^{-4}$ above which 
spherical shocks of this Mach number are CR dominated 
according to Berezhko et al.~(\cite{BeKsYe95}). 

\item{}
Due to computational limitations of using a Bohm type diffusion model, 
we have integrated our models
until the maximum momentum reaches about $(p_{\rm max}/m_{\rm p}c) \sim 1$.
For the $M=30$ shock model, the energy flux in the total CR 
distribution was about $18\%\pm5\%$ of the energy flux in the 
thermal plasma and shocks didn't become CR dominated and 
smoothed completely by the end of our simulations. 
For the $M=2.24$ shock model, 
the acceleration efficiency is lower by a factor of two 
compared to the high Mach shock because of the smaller velocity jump
across the shock. 

\item{}
Just above the injection pool, the distribution function changes sharply
from a Maxwell distribution to an approximate power--law whose index is 
close to the test--particle slope.
We estimated this critical momentum as $\pinj \sim (2.2-2.3) \cdot \pth$
where $\pth = 2\sqrt{ m_{\rm p} \boltz T}$. 
This determines the number of particles in the injection pool by
$f(\pinj) \propto \exp ( - \pinj^2 / 2 m_{\rm p} \boltz T ) $.
For strong shocks this translates into a distribution function at injection energies of
$g(\pinj) \sim (1/100 - 1/200) g(\pth)$.
 
\end{enumerate}

While the weak shock model of $M=2.24$ reaches a steady--state,
the strong shock model of $M=30$ has not reached a steady--state
by the end of our simulation. 
We expect for the strong shock that
the CR pressure continues to increase and the shock becomes CR
dominated, leading to the greater total velocity jump and more
efficient acceleration.  
In realistic shocks such as SNRs, however, escaping particles due to 
non--planar geometry or lack of scattering at high momentum are likely
to become important. 
To resolve this non--linear evolution, much longer physical time scales 
have to be simulated, until
CRs reach energies where escape is likely to be important.
The key problem here is the range in configuration and momentum space 
that has to be computed. Our method uses a grid with uniform cells
in configuration space, chosen fine enough to capture the evolution of
$g(x,p)$ at near--thermal momenta where 
the diffusion coefficient is proportional to $p^2$ (Bohm diffusion). 
This leads to a computationally extremely expensive calculation, 
especially because the grid has to be large enough to contain the 
diffusion length scale of the highest momentum CRs. 
The problem can be solved on a much larger time scale by using an
adaptive mesh refinement (AMR) code with the shock tracking techniques 
(Kang \& Jones~\cite{KaJo99}). 
In the near future we plan to incorporate the injection model 
presented here into the powerful shock tracking AMR--code, 
to calculate the evolution of the phase--space distribution of 
the plasma during different phases of SNRs. 
This would allow us to investigate 
with a plasma--physical based injection model
how the slowly growing cosmic--ray pressure
at a strong shock eventually modifies the shock structure. 
A strong modification will cause the velocity jump across the
subshock to decrease and the distribution function of the
suprathermal particles to steepens.
This might have further back reaction on the injection efficiency.
Also the CR distribution will deviate from a simple power--law.
For a calculation up to the highest energy CRs, also the spherical
geometry of a SNR should be taken into account.
Such an approach could lead to a consistent calculation of the complete
phase--space distribution at quasi--parallel shocks, and should be a promising
step towards a calculation of the overall efficiency of SNRs in producing 
CRs during their evolution.

For oblique shocks, the injection efficiencies calculated here for a parallel
shock should define an upper limit,
because the statistical probability of a particle 
to cross the shock from downstream to upstream decreases with the
intersection velocity of magnetic field and shock front. This kinematical
effect was investigated by Baring et al.~(\cite{BaElJo93}) with the use
of Monte--Carlo simulations. However, in the model we have incorporated here,
the injection is already suppressed strongly (compared to the purely 
kinematical model) by the reduced transparency
of the plasma due to the high amplitude Alfv\'en waves. We point out, that
for oblique shocks, the filtering due to Alfv\'en waves may be reduced
due to the decreased downstream amplification of the wave amplitude.
This would allow lower energy particles to be injected, and the kinematical 
effect could be partly balanced. As a result, we speculate that
the dependence on the obliquity
might be significantly weaker than calculated by 
Baring et al.~(\cite{BaElJo93}). Resolution of that important question
must await more complete  understanding of the injection physics.

In summary, we have shown that the process of particle acceleration under
consideration of a plasma physical injection model underlies a rather effective
self--regulation.
Apart from the direct particle--wave interaction described by the
injection model itself,
also the energetic feedback of the energy transfer between thermal plasma
and cosmic--rays keeps the fraction of particles in the non--thermal 
distribution at roughly $10^{-3}$ of the particles swept through the shock. 
These self--regulation mechanisms lead to a quite stable injection efficiency,
which depends weakly on the initial conditions. 
\vspace*{-0.3cm}
\section{Acknowledgments}
We are grateful to M.A. Malkov for very helpful discussions. This work was 
supported by the University of Minnesota Supercomputing Institute, by NSF grant
AST-9619438 and by NASA grant NAG5-5055.
HK's work was supported by Korea Research Foundation Grant
(KRF99-015-DI0114).
\vspace*{-0.3cm}

\end{document}